# Nanoscaled magnon transistor based on stimulated three-magnon splitting

Xu Ge[1], Roman Verba[2], Philipp Pirro[3], Andrii V. Chumak[4], Qi Wang[1,*]

[1] *School of Physics, Huazhong University of Science and Technology, Wuhan, China*
[2] *Institute of Magnetism, Kyiv, Ukraine*
[3] *Fachbereich Physik and Landesforschungszentrum OPTIMAS, Rheinland-Pfälzische Technische Universität Kaiserslautern-Landau, Kaiserslautern, Germany*
[4] *Faculty of Physics, University of Vienna, Vienna, Austria*

**Abstract**

Magnonics is a rapidly growing field, attracting much attention for its potential applications in data transport and processing. Many individual magnonic devices have been proposed and realized in laboratories. However, an integrated magnonic circuit with several separate magnonic elements has yet not been reported due to the lack of a magnonic amplifier to compensate for transport and processing losses. The magnon transistor reported in [Nat. Commun. **5**, 4700, (2014)] could only achieve a gain of 1.8, which is insufficient in many practical cases. Here, we use the stimulated three-magnon splitting phenomenon to numerically propose a concept of magnon transistor in which the energy of the gate magnons at 14.6 GHz is directly pumped into the energy of the source magnons at 4.2 GHz, thus achieving the gain of 9. The structure is based on the 100 nm wide YIG nano-waveguides, a directional coupler is used to mix the source and gate magnons, and a dual-band magnonic crystal is used to filter out the gate and idler magnons at 10.4 GHz frequency. The magnon transistor preserves the phase of the signal and the design allows integration into a magnon circuit.

Spin waves (magnons), having low intrinsic losses, high-frequency range (gigahertz to terahertz), and short wavelengths (down to several nanometers), are promising candidates for data transport and processing devices [1-5]. In contrast to sound and light waves, spin waves exhibit stronger and more diverse intrinsic nonlinear phenomena [6-9], making it easier to construct all-magnonic integrated circuits, in which the magnons are controlled by the magnons themselves without any intermediate conversion to electric currents. In the last decade, several individual magnonic devices have been proposed such as spin-wave logic gate [10,11], magnon transistor [12,13], majority gate [14,15], magnon valve [16,17], and directional coupler [18,19]. However, the realization of an integrated magnonic network [20,21] with several separate magnonic devices is still a challenge. The main reason is that spin wave amplitude decreases after passing through the upper-level device due to magnetic damping and could not reach the nonlinear threshold of the next-level device. The key element to solve this problem is the magnonic amplifier, which compensates the loss and brings the spin-wave amplitude back to the initial state.

In recent years, researchers have explored effective ways to amplify propagating spin waves. One of the methods is based on the reduction of magnetic damping, which enhances spin-wave signals by spin transfer torque and spin orbit torque generated by a DC current [22,23]. Recently, Merbouche *et al.* [24] reported a true amplification of spin waves based on spin currents. In this research, the direction of the external field and the composition of the materials were precisely tuned to avoid the occurrence of nonlinear magnon scattering and auto-oscillations. Another common approach is parallel parametric pumping, in which one microwave photon splits into two magnons at half the frequency under the conservation of energy and momentum to amplify the propagating spin waves [25-27]. Very recently, Breitbach et al. [28] proposed a spin-wave amplifier based on rapid cooling, which is a purely thermal effect. The magnon system is brought into a state of local disequilibrium with an excess of magnons. A propagating spin-wave packet reaching this region stimulates the subsequent redistribution process, and is in turn amplified.

---
[*]Author to whom correspondence should be addressed: williamqiwang@hust.edu.cn

In modern CMOS electronics, transistors are used to amplify electrical signals. The magnon transistor reported in [12] was designed to suppress one magnon flux by another in order to perform all-magnon logic operations. Since the suppression efficiency was very high, it was shown that using a special interferometric scheme, the transistor could also be used for amplification, but with a rather small gain of 1.8. Another magnon transistor concept suitable for direct amplification of the magnon fluxes is needed.

In this letter, we propose a magnon transistor concept based on stimulated three-magnon splitting and validate it using micromagnetic simulations. First, we study three-magnon scattering in a nanoscale straight magnonic waveguide with different external field directions. A pronounced three-magnon scattering is observed, in which one gate (pump) magnon (14.6 GHz) splits into two magnons (10.4 GHz and 4.2 GHz) according to the laws of energy and momentum conservation. It is found that the additional injection of the source magnons at the frequency of 4.2 GHz leads to a drastic enhancement of the splitting of the gate magnons, a phenomenon we refer to as stimulated magnon splitting. Based on this mechanism, we propose a magnon transistor design that employs a directional coupler to mix gate and source magnons. In addition, a dual-band magnon crystal is used to filter out the gate (14.6 GHz) and idler (10.4 GHz) magnons from the transistor drain. The transistor allows the drain magnon density to be 9 times higher than the source magnon density.

As shown in Fig. 1(a), we consider a yttrium iron garnet (YIG) waveguide of length $l = 20$ μm, width $w = 100$ nm, and thickness $t = 50$ nm. To investigate the effects of three-magnon splitting, micromagnetic simulations are preformed using Mumax$^3$ [29] with the following parameters of YIG [30]: saturation magnetization $M_s = 1.4 \times 10^5$ A/m, exchange constant $A = 3.5 \times 10^{-12}$ J/m, and damping coefficient $\alpha = 2 \times 10^{-4}$. An external bias magnetic field $B_{ext} = 100$ mT is applied in the $xy$ plane forming an angle $\varphi_H$ with the negative $x$-axis, as illustrated in Fig. 1(a). The maximum splitting efficiency is obtained at approximately $\varphi_H = 60°$, which corresponds to $\varphi_M = 45°$ (the angle between the magnetization direction and the negative $x$-axis). It is a known feature of the three-magnon interaction in an effectively one-dimensional system [31], which is additionally explained in the supplementary materials. The magnetization direction is not aligned with the direction of the external field due to the nonnegligible demagnetic field along the width direction in the nanoscale waveguide. In the following studies, the external field is fixed at the optimal value $\varphi_H = 60°$. The prorogating spin wave in the waveguides is excited by an alternating field as a sinusoidal function of time: $\mathbf{h}_{rf} = b\sin(2\pi ft)\mathbf{e}_z$ with the oscillation amplitude $b$ and the excitation frequency $f$. The applied field is in the center of waveguide over an area of 30 nm in length [blue area shown in Fig. 1(a)]. To avoid spin wave reflection, the damping coefficient has an exponential increase to 0.5 at both ends of the waveguide [32]. In addition, all the simulations are performed with room temperature $T = 300$ K. We recorded the time-dependent magnetization data from $t = 0$ ns to 50 ns across the entire waveguide and then obtained the spin wave dispersion curves by using two-dimensional fast

Fourier transform [33,34].

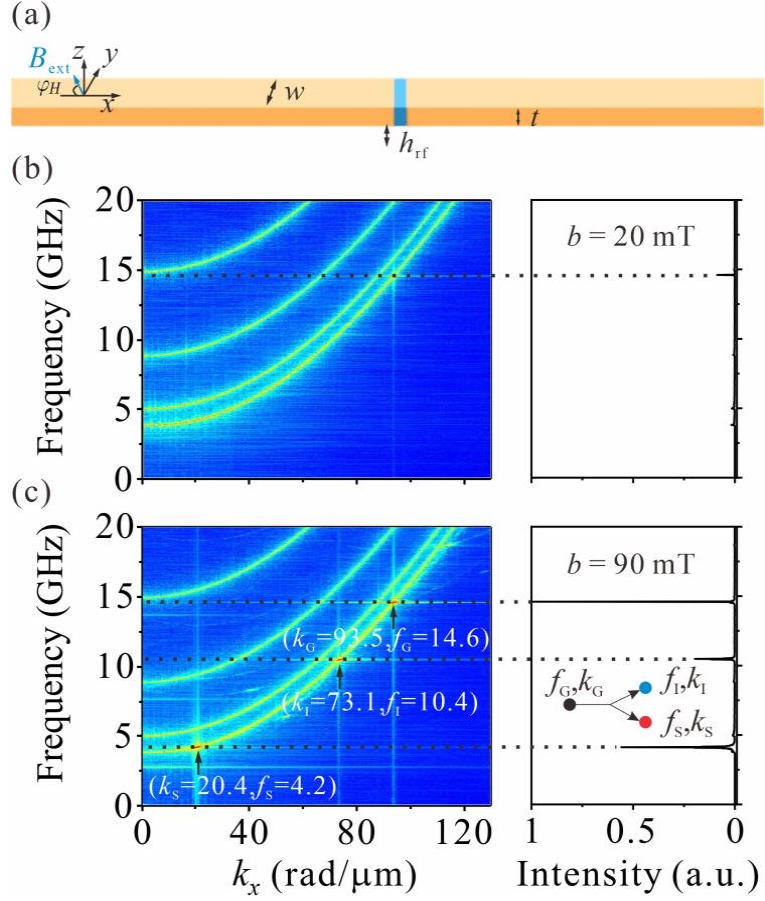

*Fig. 1. (a) A sketch of spin-wave excitation system: a 30-nm wide excitation region is used to generate oscillation magnetic field $\mathbf{h}_{rf}$, and the in-plane magnetic field is applied at $\varphi_H = 60°$. Dispersion curve of the propagating spin waves and spin-wave intensity as a function of frequency for the different amplitudes of excitation field are shown in (b) b = 20 mT and (c) b = 90 mT.*

In the current study, the excitation frequency $f_G$ is considered to be 14.6 GHz. The left part of Fig. 1(b) and 1(c) shows the dispersive curves corresponding to different excitation field amplitudes. At $b$ = 20 mT, the dispersive curve exhibits only one weak bright spot [Fig. 1(b) left], and it can also be seen that there is only one weak peak in the spin-wave intensity spectrum, corresponding to the directly pumped magnon of frequency 14.6 GHz [Fig. 1(b) right]. When the amplitude $b$ is increased to 90 mT, two other stronger bright spots at lower frequencies appear in the dispersion curve [Fig. 1(c) left]. These peaks, observed at 14.6, 10.4, and 4.2 GHz, correspond to three distinct magnons: the gate magnon $f_G$ ($k_G$ = 93.5 rad/μm) and the scattering idler magnons $f_I$ ($k_I$ = 73.1 rad/μm) and source magnons $f_S$ ($k_S$ = 20.4 rad/μm), as shown in the right of Fig. 1(c). This observed process adheres to the energy-momentum conservation laws, satisfying $f_G = f_I + f_S$, $k_G = k_I + k_S$. This phenomenon is nothing else than three-magnon splitting [31,35,36].

Now we consider stimulated three-magnon scattering, by exciting both the gate and one of the splitting (source) magnons. Namely, we considered the excited field as $\mathbf{h}_{rf} = b_S\sin(2\pi f_S t)\mathbf{e}_z + b_G\sin(2\pi f_G t)\mathbf{e}_z$ with one more excitation frequency $f_S$ = 4.2 GHz, and collected the spin-wave intensity of drain magnons having frequency equal to the frequency of the source magnons $f_D$ =

$f_S$ = 4.2 GHz for the varying amplitude of excitation field $b_G$ in the range from 1 mT to 130 mT. As shown in Fig. 2, the spontaneous three-magnon splitting process appears when $b_G$ is greater than 30 mT, see black line (rigorously speaking, it is not absolute three-wave instability, but a convective one [37, 38]). The red and blue lines correspond to the stimulated process, and were obtained by applying two excitation frequencies of $f_G$ = 14.6 GHz and $f_S$= 4.2 GHz. The spin-wave intensity spectra of the red and blue lines are obtained by subtracting the energy of the directly excited magnons at a frequency of 4.2 GHz, and thus show the same ground intensity as the black line. First, we see, that stimulated process takes place at a lower gate power than the spontaneous one and, in fact, does not demonstrate any threshold. This feature is expected for three-wave processes and, for magnon system, in particular, was demonstrated in [36] by inspecting idler magnon density.

For the same excitation amplitude $b_G$, the red and blue lines show stronger spin-wave intensity in contrast to the black line. This indicates that the existing magnon of frequency $f_S$ = 4.2 GHz can stimulate not only the appearance of idler magnons (as shown in [36]), but also increases the population of the source magnons itself, and the scattering efficiency depends on the source magnon density. In addition, the stronger density of scattering magnon is achieved when the amplitude $b_S$ of $f_S$ = 4.2 GHz is increased. This enhancement of the three-magnon splitting by introducing one of the splitting magnons can be used for the amplification of propagation spin waves.

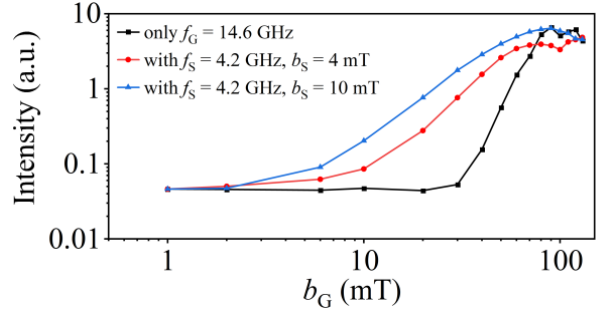

*FiG. 2. Intensity of the scattering spin waves at frequency of $f_D$ = 4.2 GHz as a function of the gate excitation field $b_G$, which is proportional to gate magnon density. The black line corresponds to the case when only the gate magnons are directly excited at $f_G$ = 14.6 GHz. The red and blue lines depict the case when gate and source spin waves are simultaneously excited by alternating field at $f_G$ = 14.6 GHz and $f_S$ = 4.2 GHz for two different densities of source magnons, driven by source excitation field $b_S$ = 4 mT (red line) and $b_S$ = 10 mT (blue line).*

Estimation of the amplification efficiency can be done in a way, similar to that used in nonlinear optics [37]. The maximum amplification rate can be calculated in the conservative approximation, i.e. neglecting the damping. As derived in the supplementary materials, the maximum amplitude of the drain magnons $A_D$ is given by:

$$A_{D,\max}^2 - A_S^2 = A_G^2 \frac{v_G}{v_S}, \quad (1)$$

where $A_S$ and $A_G$ are initial amplitudes of source and gate magnons at $f_S$ and $f_G$, respectively, while $v_S$ and $v_G$ are their group velocities; amplitudes are defined as standard canonical amplitudes [31]. This linear dependence of the gain on the gate power $\left(A_{D,\max}^2 - A_S^2\right) \sim b_G^2$ is nicely reproduced in the simulations; only at high gate powers, above $b_G$ > 50 mT, the gain saturates due to the influence of other nonlinear effects, in particular, self- and cross- nonlinear frequency shift. We see, that the amplification rate $A_D/A_S$ is larger for smaller source amplitude $A_S$, since the total power, which can be transferred to the source wave, is limited by the gate magnon power. Also, amplification efficiency increases with the ratio $v_G/v_S$. In the case of spin waves in waveguides, higher-frequency (gate) magnon almost always has larger group velocity than lower-frequency (source) magnon, improving the amplification mechanism (In our case, the ratio $v_G/v_S$, estimated to be approximately 6.8, is derived from the dispersive curve depicted in Fig. 1(c)). Of course, damping decreases the amplification efficiency and introduces a dependence on the transferred power on the initial power of source magnons. Detailed consideration of this process lies beyond the scope of this work.

A magnon transistor is designed as shown in Fig. 3(a) consisting of one magnonic directional coupler [39] and a dual-band magnonic crystal with different periods [40]. The directional coupler [41] is employed for combining spin wave signals with different frequencies from two separated waveguides into one waveguide via frequency-dependent dipolar coupling strength [19,39]. The directional coupler is carefully designed with the following geometric dimensions: the length of coupled waveguides ($L_1$) is 390 nm, the angle between waveguides ($\Phi$) is 10°, the gap between coupled waveguides ($\delta$) is 10 nm, and the edge-to-edge distance ($d_1$) is 200 nm, to make sure that gate magnons can pass through it without significant energy loss, and the source magnons will be completely guided from the bottom waveguide to the top one as shown by the black and red arrows in Fig. 3(a). Once the source magnon is coupled into the top waveguide and mixed with the gate magnon, the three-magnon splitting is enhanced and the gate magnon efficiently splits into an idler magnon (10.4 GHz) and a drain magnon with the frequency of 4.2 GHz resulting in an amplification of the source magnon.

Figure 3(b) shows the working principle of the magnon transistor. When only the gate magnon ($f_G$ = 14.6 GHz, $b_G$ = 40 mT) is applied, the two-dimensional colormap shows the weak magnon density of 4.2 GHz due to the low three magnon splitting efficiency [top of Fig. 3(b)]. The middle panel of Fig. 3(b) shows the case of only the source magnon is injected from the bottom waveguide. It clearly shows that all the source magnons are guided from the bottom waveguide to top waveguide (transistor's drain) by the directional coupler as expected. Once both gate and source magnons are simultaneously excited, the drain magnon density at 4.2 GHz is dramatically increased due to the stimulation of the three magnon splitting [bottom of Fig. 3(b)]. In order to get only the flux of drain magnons at the output, a specially designed dual-band magnonic crystal, in the form of two in-line connected magnonic crystals of different periodicities, is used to filter out the gate and idler magnons produced by three magnons splitting [42]. The design principle of magnonic crystals is discussed in detail in supplementary materials. Figure 3(c) shows the magnetization oscillations spectra extracted from the end of the top waveguide as marked by the red dashed rectangular regions in Fig. 3(a). It shows two properties: (1) The source magnon has a gain factor of 9. (2) The gate magnon and idler magnon have been efficiently suppressed by the dual-band magnonic crystal. Importantly, that such a large gain is observed for moderate source magnon power, which is required for logic applications – amplification of nonlinear spin waves is much more nontrivial task than of small-amplitude ones [43]. Furthermore, the amplification is not sensitive to the relative phase between the gate and source magnons (see the supplementary materials). Therefore, the output signal can be directly used to connect the next logic gate without any phase modulations and the proposed magnonic transistor is suitable for further integration.

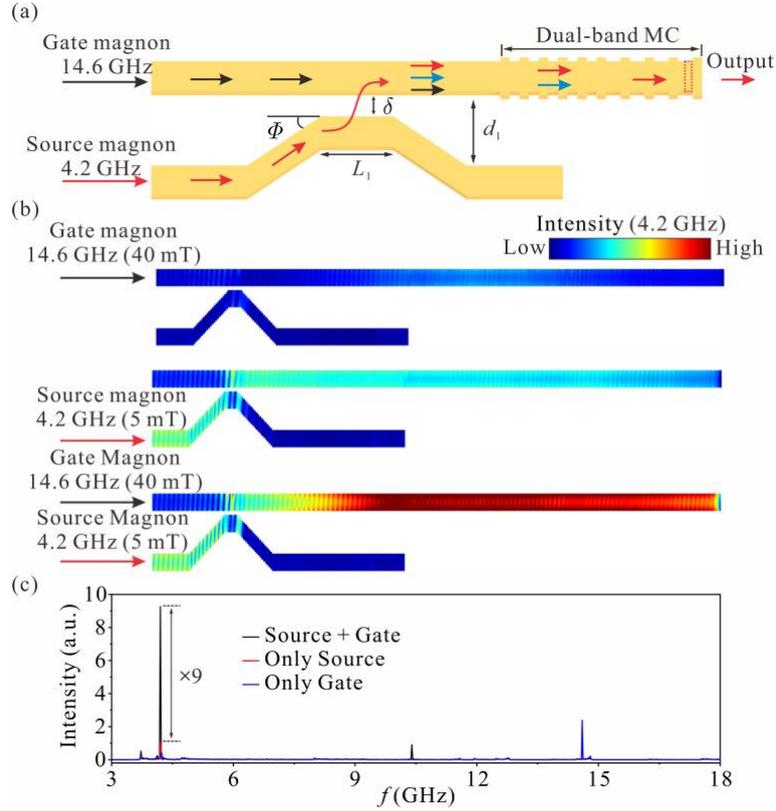

*Fig. 3. (a) Schematic of magnon transistor designed with a directional coupler with dual-band magnonic crystal. (b) The intensity distributions of spin waves at 4.2 GHz and (c) the magnetization oscillation spectrum extracted at the end of the top waveguide marked by red dashed rectangular regions for different cases.*

In summary, we numerically demonstrate magnon transistor based on the phenomenon of stimulated three-magnon splitting within a nano-scaled waveguide with an in-plane inclined static external field. The three-magnon splitting efficiency of the gate magnons is substantially enhanced by directly introducing one of the signal magnons. A magnon directional coupler is used to mix the gate and source magnons, and a dual-band magnon crystal is used to filter the idler and gate magnons from the drain of the transistor. The transistor has a gain of 9, measured as the ratio of the drain magnon density normalized to the source magnon density. The phase of the drain magnons depends only on the phase of the source magnons, and the design of the transistor with separate source, gate and drain magnon conduits makes it suitable for further integration into a complex magnonic network. This device represents a step forward in the amplification of propagating spin waves in nanoscale waveguides, offering promising avenues for the advancement of spintronic applications.

This work was supported from the National Key Research and Development Program of China (Grant No. 2023YFA1406600) and National Natural Science Foundation of China, the startup grant of Huazhong University of Science and Technology (Grant No. 3034012104). R.V. Acknowledge support of the MES of Ukraine.

# Supplementary Materials for

# "Nanoscaled magnon transistor based on stimulated three-magnon splitting"

Xu Ge[1], Roman Verba[2], Philipp Pirro[3], Andrii V. Chumak[4], Qi Wang[1]

[1] *School of Physics, Huazhong University of Science and Technology, Wuhan, China*
[2] *Institute of Magnetism, Kyiv, Ukraine*
[3] *Fachbereich Physik and Landesforschungszentrum OPTIMAS, Rheinland-Pfälzische Technische Universität Kaiserslautern-Landau, Kaiserslautern, Germany*
[4] *Faculty of Physics, University of Vienna, Vienna, Austria*

**S1. Identification the inclination angle $\varphi_H$ based on the scattering intensity**

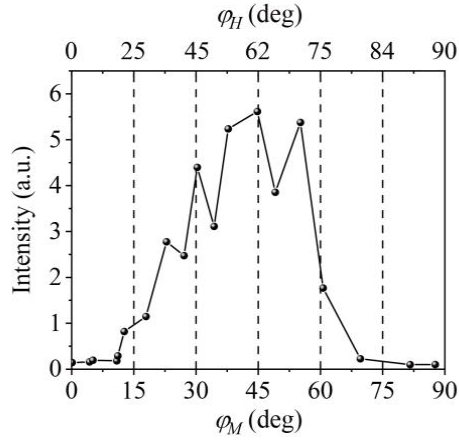

*Figure S1. The scattering spin-wave intensity as function of the angle $\varphi_M$ the angel between the magnetization direction and negative x-axis (bottom transverse axis).*

The three-magnon Hamiltonian, governing considered process, is derived as

$$H^{(3)} = \sum_{123} \left( V_{12,3} c_1 c_2 c_3^* + \text{c.c.} \right) \Delta(\mathbf{k}_1 + \mathbf{k}_2 - \mathbf{k}_3) \tag{S1}$$

where $\mathbf{k}_i = k_i \mathbf{e}_x$ is spin-wave wave vector (1, 2, and 3 denote the three magnons), directed along the waveguide (x direction). Spin-wave amplitudes $c_i$ are defined as usual canonic variables [1], i.e. have an order of dimensionless dynamic magnetization. In our case, magnons 1, 2, and 3 are source, idle and gate magnons, respectively.

The three-magnon coefficient $V_{12,3}$ could be found using scalar [1] or vectorial [2] spin-wave Hamiltonian formalism. General expression is quite cumbersome and is not presented here. However, using a simplified expression for the dynamic demagnetization tensor equal to one of a film, and neglecting minor ellipticity-related term, three-magnon coefficient is derived as

$$V_{12,3} \approx \frac{\omega_M \cos\varphi_M \sin\varphi_M}{2\sqrt{2}} \left( f(k_1 t) + f(k_2 t) \right), \tag{S2}$$

where $\omega_M = \gamma \mu_0 M_S$, with $\gamma$ representing the gyromagnetic ratio, $\mu_0$ as the vacuum permeability, and $M_S$ as the saturation magnetization. Additionally, $f(x) = 1 - (1 - \exp[-|x|])/|x|$ is the so-called thin-film function. Consequently, the calculated value of $\varphi_M = 45°$ corresponds to the maximum strength of three-magnon splitting determined by the above theoretical equation of the scattering coefficient [Eq. (S2)]. Furthermore, as seen in Fig. S1, we extracted the magnetization direction angle $\varphi_M$ and the intensity of lower frequency scattering magnons by varying the direction of the in-plane external field from $\varphi_H = 0°$ to $\varphi_H = 90°$. The strongest peak is observed at around $\varphi_M = 45°$, which is consistent with the above theoretical analysis, in turn, the corresponding angle $\varphi_H$ is found to be around 60° (Fig. S1), and hence, a tilt angle $\varphi_H = 60°$ of the external field is utilized to investigate the magnon scattering in the main text.

**S2 Spin-wave amplification by three-magnon scattering**

Using well-developed approach from nonlinear optics [3], one can derive equations for spatial-temporal evolution of spin waves envelope amplitudes $a_i = a_i(x, t)$:

$$\begin{aligned}
(\partial_t + v_1\partial_x + \Gamma_1)a_1 &= -2iVa_2^*a_3, \\
(\partial_t + v_2\partial_x + \Gamma_2)a_2 &= -2iVa_1^*a_3, \\
(\partial_t + v_3\partial_x + \Gamma_3)a_3 &= -2iVa_1a_2.
\end{aligned} \quad (S3)$$

Here $v_i$ is spin-wave group velocity, $\Gamma_i$ the damping rate, $V \equiv |V_{1,2,3}|$, and equations are written for the resonant case, i.e. when $f_3 = f_1 + f_2$.

Eq. (S3) does not allow for exact analytical solution. Let's first look on the amplification conditions. For this, assume the pumping wave amplitude $a_3$ to be constant, signal wave initial value $a_1(x = 0) = A_1$ and absent idler wave $a_2(x = 0) = 0$. Also, stationary regime ($\partial_t \to 0$) is considered. Then, solution of the first two equations of Eq. (S3) is a simple sum of exponents $a_1 \sim c_1 e^{\kappa_1 x} + c_2 e^{\kappa_2 x}$, where

$$\kappa_{1,2} = \frac{1}{v_1 v_2}\left(-(v_1\Gamma_2 + v_2\Gamma_2) \pm \sqrt{(v_1\Gamma_2 + v_2\Gamma_2)^2 + 4v_1v_2(4|VA_3|^2 - \Gamma_1\Gamma_2)}\right). \quad (S4)$$

It is clear, that if $|2VA_3| > \sqrt{\Gamma_1\Gamma_2}$, one of the exponent becomes positive, $\kappa_1 > 0$, meaning amplification of signal wave. This condition is independent of initial amplitude of the signal spin-wave and is common for any parametric pumping process.

To estimate amplification rate, we look on conservative approximation of Eq. (S3), i.e. neglect the damping. Then, introducing new variables $b_i = 2Va_i/\sqrt{v_jv_l}$, $j, l \neq i$, this system is reduced to standard form:

$$\begin{aligned}
\partial_x b_1 &= -ib_2^*b_3, \\
\partial_x b_2 &= -ib_1^*b_3, \\
\partial_x b_3 &= -ib_1b_2.
\end{aligned} \quad (S5)$$

Exact solution of this system is expressed via elliptic functions and could be found in [3]. Here we note only that this system possesses three integrals of motion (strictly speaking, only two of which are independent), called the Manley-Rowe relations: $|b_1(x)|^2 - |b_2(x)|^2 = $ const, $|b_1(x)|^2 + |b_3(x)|^2 = $ const, and $|b_2(x)|^2 + |b_3(x)|^2 = $ const. It is clear, that the powers of signal and idler wave change synchronously – either both increase taking energy from the pumping, either both decrease when the energy flows back to the pumping wave. Also, the power of signal wave cannot be smaller than the the initial power, as the initial power of idler wave is zero. Thus, signal wave is always amplified in the conservative approximation. The maximal power is $|b_1|^2_{max} = |b_1(0)|^2 + |b_3(0)|^2$. Returning to the initial variables, we get the gain

$$|a_1|^2_{max} - A_1^2 = A_3^2 \frac{v_3}{v_1}, \quad (S6)$$

which is presented in the main text [Eq. (1)] accounting for the notation of gate, source and drain spin-wave amplitudes.

### S3. Frequency filtering with nanostrip magnonic crystals

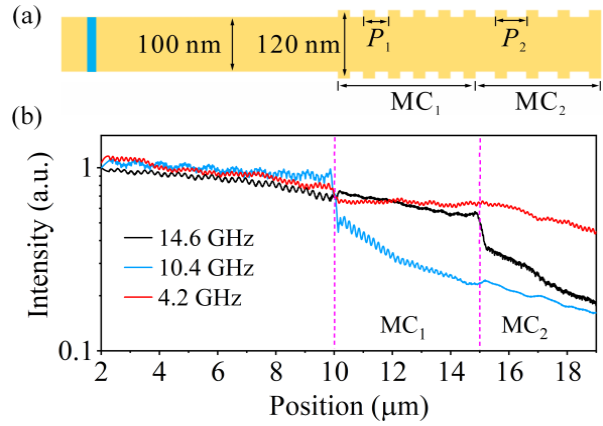

*Figure S2. (a) Schematic of magnonic filter: spin waves are generated in the blue area of magnonic waveguide, and filtered out in the dual-band magnonic crystals (labeled as MC$_1$ and MC$_2$) region. (b) Intensities of propagating spin waves at different positions along the waveguide for the indicated frequencies of 14.6, 10.4, and 4.2 GHz.*

The propagation characteristics of spin waves can be controlled through spatially periodic modulation in magnonic crystals, effectively serving as spin-wave filters due to band gaps [4,5]. As seen in Fig. S2(a) and Fig. 3(a) of the main text, we demonstrate the dual-band magnonic filter composed of two parts. One is the 100-nm wide spin-wave propagation waveguide in the left and the other parts are two magnonic crystals (MC$_1$ and MC$_2$) with periodicity $P_1 = 40$ nm and $P_2 = 50$ nm, respectively. Fig. S2(b) presents the calculated intensities of spin waves propagating along the waveguide. After traveling through MC$_1$ and MC$_2$, the spin-wave intensities with frequencies of 14.6 and 10.4 GHz (black and blue lines) are reduced by more than 70%, while spin wave with frequency of 4.2 GHz (red line) can propagate through the two magnonic crystals without additional loss.

### S4. Dependence of the output spin-wave intensity on the relative phase between the pumping and source magnons

For construction of an integrated magnonic circuit, it is hard to accurately control the relative phase between the pumping and source magnons. To elucidate the relationship between the output spin-wave intensity and the relative phase between the pumping and source magnons, we introduced a variable, denoted as $\phi$, representing the relative phase between two magnons. Subsequently, we recorded the output spin-wave intensity at 4.2 GHz for various $\phi$ values spanning from 0° to 360°. The results, depicted in Fig. S3, indicate that the output spin-wave intensity exhibits relatively low sensitivity to alterations in the relative phase between the gate and source magnons. The results suggest that the magnonic amplifier based on three-magnon scattering has potential to help to construct an integrated magnonic circuits.

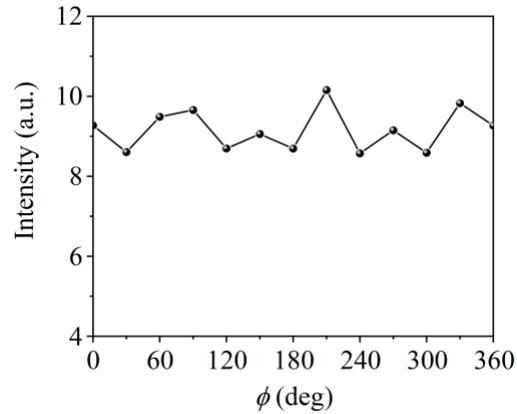

*Figure S3 The output spin-wave intensity with frequency of 4.2 GHz as function of the varied relative phase between the gate and source magnons*